\documentclass{article}
\usepackage[utf8]{inputenc}
\usepackage{amsmath}
\usepackage{amsfonts}
\usepackage{braket}
\usepackage{graphicx}
\usepackage{physics}
\usepackage{xcolor}
\usepackage{multirow}
\usepackage{adjustbox}
\usepackage[a4paper, total={6in, 8in}]{geometry}

\title{\textbf{Interferometric weak value of polarization observable and differential Jones matrix algebra}}
\author{Niladri Modak$^{1,*}$, Sayantan Das$^{1}$, Priyanuj Bordoloi$^{1}$, Nirmalya Ghosh$^{1}$\\
$^{1}$\textit{\small Department of Physical Sciences},\\ \textit{\small Indian Institute of Science Education and Research Kolkata,}\\ \textit{\small Mohanpur, India- 741246}}

\date{\footnotesize$^{*}$nm16ip018@iiserkol.ac.in}

\begin{document}

\maketitle
\begin{abstract}
    \noindent 
    The quantification of simultaneously present weak polarization anisotropy effects are of practical interest from polarimetric and metrological perspective. Recently, in Phys. Rev. A 103, 053518, we experimentally demonstrated a classical analog of post-selected quantum weak measurement through optical interferometry to amplify all possible weak polarization anisotropy effects individually. Here, we propose an extension of this interferometric framework to quantify simultaneously present polarization anisotropy effects. Moreover, a clear correspondence of differential Jones matrix approach with the present scheme is indicated. The proposed scheme enables the measurement of differential Jones matrices through characteristic Stokes vector elements. Our proposal leads to a new class of polarimeter for experimental detection of differential Jones matrix of non-depolarizing anisotropic medium exhibiting simultaneous multiple polarimetric effects of tiny magnitude.
\end{abstract}
\section{Introduction}
The difference in refractive indices between orthogonal polarization states gives rise to the polarization anisotropy effects namely diattenuation and retardance \cite{collett1993polarized,gupta2015wave,perez2016polarized}. Generally, such anisotropy effects of a sample are characerized by standard polarimetry techniques by measuring Stokes vector and calcualting the Mueller matrix \cite{collett1993polarized,gupta2015wave,perez2016polarized,ghosh2011tissue,azzam2016stokes}. Although there had been significant developments of this field, there is always a continuous effort to develop fundamental understandings which in turn, has generated numerous applications spanning all branches of science \cite{collett1993polarized,gupta2015wave,perez2016polarized,ghosh2011tissue,azzam2016stokes,bliokh2008geometrodynamics,bliokh2015quantum,aiello2015transverse,lin2014dielectric,gadsden1979detection,chrysostomou2007circular,ren2016nanotwin,gurjar2001imaging,fasolka2003measuring,belotelov2011enhanced,salvail2013full,rhee2015chiroptical,luo2019simultaneously,xu2019multifunctional,pryde2005measurement,iinuma2011weak,de2014ultrasmall,zhu2019single}. Such measurement of polarization properties had been potentially applied in numerous areas from understanding chemical structures of bio-chemical compounds \cite{collett1993polarized,gupta2015wave,perez2016polarized,ghosh2011tissue,azzam2016stokes,bliokh2008geometrodynamics,bliokh2015quantum,fasolka2003measuring}, biomedical imaging \cite{gurjar2001imaging}, microscopy \cite{rizzo2005high}, spectroscopy \cite{ray2017polarization}, and so on \cite{chrysostomou2007circular,ren2016nanotwin,fasolka2003measuring}. However, these methods suffer from being extremely tedious, and the fact that the anisotropy effects, often very small, were very hard to probe. We recently came up with an interferometric protocol to characterize such small anisotropy effects using the weak value amplification (WVA) technique realized in a traditional classical Mach-Zehnder interferometer \cite{modak2021generalized}.
\par
Since its discovery \cite{aharonov1988result}, quantum weak measurement and WVA achieved considerable attention, both from the perspective of theoretical foundations of physics and potential multi-disciplinary applications \cite{ritchie1991realization,hosten2008observation,magana2014amplification,dixon2009ultrasensitive,xu2013phase,asano2016anomalous,PhysRevA.89.053816,nechayev2018weak,shomroni2013demonstration,singh2018tunable,lundeen2011direct,lundeen2009experimental,palacios2010experimental,kocsis2011observing,luo2019simultaneously,xu2019multifunctional,pryde2005measurement,iinuma2011weak,de2014ultrasmall}. `Weak value' (WV) appears as a consequence of pre and post-selection (PPS) aided weak measurement \cite{aharonov1988result,duck1989sense,dressel2014colloquium,kofman2012nonperturbative,ritchie1991realization}. WV can be complex and lie far outside the eigenvalue spectrum of the measuring operator depending on the relative overlap between pre and post-selected states, proving itself a competent candidate for enhancing a tiny signal and making it detectable \cite{aharonov1988result,duck1989sense,dressel2014colloquium,kofman2012nonperturbative,ritchie1991realization}. The optical analog of WVA has been widely and extensively used to amplify small signals in the context of classical and quantum optics \cite{hosten2008observation,magana2014amplification,dixon2009ultrasensitive,xu2013phase,asano2016anomalous,PhysRevA.89.053816,nechayev2018weak,shomroni2013demonstration,singh2018tunable,lundeen2011direct,lundeen2009experimental,palacios2010experimental,kocsis2011observing,luo2019simultaneously,xu2019multifunctional,pryde2005measurement,iinuma2011weak,de2014ultrasmall}. In classical optics, WVA has successfully been utilized to detect optical spin-Hall effect \cite{hosten2008observation}, sub-wavelength ordered optical beam shifts \cite{hosten2008observation,pal2019experimental,dixon2009ultrasensitive}, small time delays \cite{asano2016anomalous}, tiny beam deflection and rotation \cite{dixon2009ultrasensitive,magana2014amplification}, frequency and phase shifts \cite{xu2013phase}, and so on \cite{PhysRevA.89.053816,nechayev2018weak,shomroni2013demonstration,singh2018tunable}.
\par
Our recently proposed interferometric weak value polarimeter \cite{modak2021generalized} was limited by its ability to characterize only one anisotropy effect at a time. However in practice, the polarization anisotropy effects do not appear individually and exclusively. In general, in a complex system, all the polarization effects can occur simultaneously and in any arbitrary order \cite{collett1993polarized,gupta2015wave,perez2016polarized,ghosh2011tissue,azzam2016stokes}. Hence, with a priory assumption of a particular order of the occurrence of the anisotropy effects, the polarimetric measurements of these elements always lead to tedious inverse analysis for obtaining the polarization response of the sample \cite{collett1993polarized,gupta2015wave,perez2016polarized,ghosh2011tissue,azzam2016stokes}. To deal with such a scenario, Jones first proposed a differential matrix, i.e. ``$\boldsymbol{N}$ matrix'' \cite{Jones:48,Arteaga:09}. Although many efforts have been put to extract the differential matrix experimentally, this is still an unresolved problem in polarization metrology.
\par
In this work, we looked for an extension of the interferometric WVA framework to simultaneously encompass multiple polarization anisotropy effects. Simultaneous weak polarization anisotropy effects introduced in one path of the interferometer provide the desirable weak coupling between the path degree of freedom and the polarization degree of freedom of light. Near destructive interference of the paths of the interferometer induces considerable enhancement of different polarization anisotropy effects reflected in the amplification of characteristic Stokes vector elements \cite{gupta2015wave}. Thus the ``weak value polarimeter" can quantify simultaneously present sample polarization anisotropy effects. Interestingly, this framework provides an experimentally feasible platform to extract the differential Jones matrices for respective polarization anisotropy effects through the characteristic Stokes vector elements.
\par
The rest of the paper is organized as follows. In section \ref{sec2}, with a brief review of \cite{modak2021generalized} the central idea to extract simultaneous polarization effects has been demonstrated taking a realistic example (in section \ref{sec2.1}). Section \ref{sec2.2} includes the concept of extracting differential Jones matrix through the experimentally measurable Stokes parameters. Section \ref{sec3} concludes the report with an outlook and importance of the proposed ``weak value polarimeter" in the good old domain of the polarimetric metrology.
\section{Interferometric WVA Invoking Simultaneous Polarization Anisotropy Effects}
\label{sec2}
Using interferometric philosophy of weak value amplification \cite{duck1989sense}, we have recently shown that through a standard Mach Zehnder interferometry setup, all possible polarization anisotropy effects, namely linear and circular retardance and diattenuation, can be quantitatively enhanced \cite{modak2021generalized}. The introduction of a weak anisotropic effect $\alpha$ (defined in \cite{modak2021generalized}) in one path of the interferometer makes the system weakly coupled with the pointer. Near destructive interference at the exit port of the interferometer results in a significant enhancement of $\alpha$. The polarization degree of freedom of light is used as the pointer, i.e., the WVA of $\alpha$ is reflected on the subsequent Stokes polarization parameters $I$, $Q$, $U$, $V$ \cite{gupta2015wave}.
\par
\begin{figure}[ht]
\centering
\includegraphics[width=0.8\linewidth]{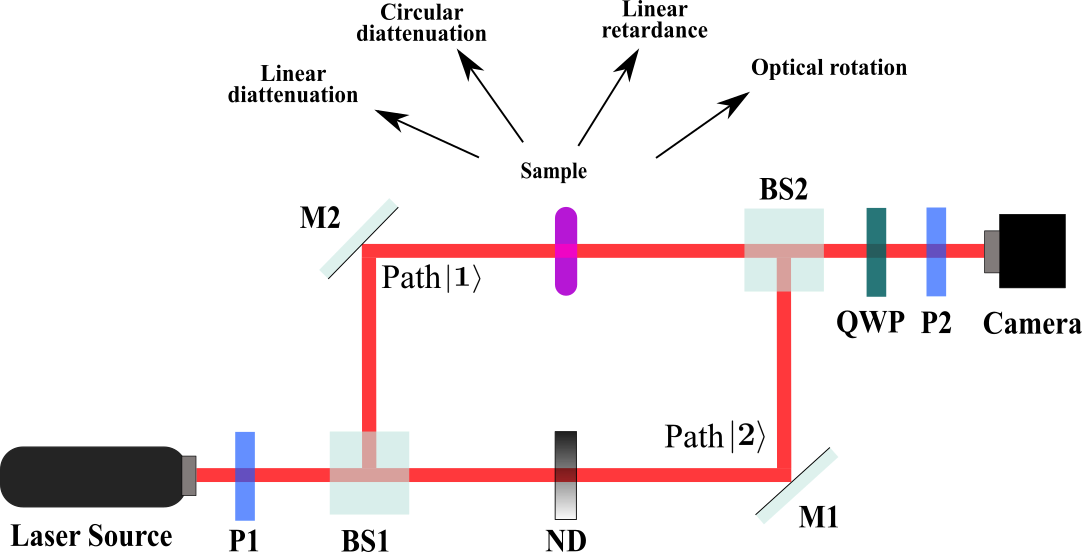}
\caption{Schematic of Mach Zehnder interferometric setup for simultaneous weak value amplification of multiple anisotropy effects. (P1,P2): polarizers, QWP: quarter waveplate, (BS1, BS2): 50:50 beam splitters, (M1, M2): mirrors, ND: variable neutral density filter. The two paths of the interferometer, path $\ket{1}$ and path $\ket{2}$ are marked. The sample in the path path $\ket{2}$ marked as purple coloured block can have all possible polarization effects as shown in the figure.}
\label{fig1}
\end{figure}

As mentioned earlier, this protocol is realized in a standard Mach Zehnder interferometer \cite{modak2021generalized, guchhait2020natural}. Table \ref{table1} demonstrates the corresponding input electric fields and subsequent characteristic Stokes vector elements for extracting different polarization anisotropy effects \cite{modak2021generalized}. In path $\ket{2}$ of the interferometer, after experiencing the anisotropy $\alpha$ (marked as purple block in Fig. \ref{fig1}), the input field $\vec{E_1}$ modifies to $\vec{E_2}$ shows enhancement of the respective Stokes vector elements listed in the third column after near destructive interference with the input beam coming from the path $\ket{1}$ of the interferometer (see Fig. \ref{fig1}). Here, near destructive interference refers to destructive interference of two fields having a phase difference $\pi$ with a small amplitude offset $\epsilon_a=\tan^{-1}({\frac{1-a}{1+a}})$ ($a$ is the ratio of amplitudes of the two fields) or interference of two fields having equal amplitude ($a = 1$)  but with a phase difference of $\pi\pm2\epsilon_p$ where $\epsilon_p$ is small phase offset. The former generates \textit{real} WVA and the latter \textit{imaginary} WVA. The quarter waveplate QWP and the linear polarizer P2 combination is used to determine the spatial variation of the relevant Stokes polarization parameters across the interference fringe obtained in the camera (see Fig. \ref{fig1}). The real WVA can be achieved by using a variable neutral density (ND) filter \cite{modak2021generalized,guchhait2020natural} which essentially varies the amplitude ratio $a$ in path $\ket{1}$ and path $\ket{2}$, thus varying the relative intensities. Variation in the relevant Stokes parameters(as in Table \ref{table1}) as a function of amplitude offset parameter $\epsilon_a$ provides the corresponding real WVA. Instead, the imaginary WVA can be probed by keeping the relative intensities in the paths same and recording the relevant Stokes parameters(as in Table \ref{table1}) for different spatial positions slightly away from the point of minimum intensity of the interference fringes which provide the phase offset parameter $\epsilon_p$ \cite{modak2021generalized,guchhait2020natural}.
\par
Even though, from Table \ref{table1} it may appear that WVA of different polarization anisotropy effects is sometimes manifested in the same Stokes vector elements, multiple polarimetry effects are perfectly discernible through real and imaginary weak value measurements and by clever choice of input polarization states.  We explain this by taking an example of detection of Stokes parameter $V/I$. There are two important issues that are needed to be noted in this regard.
\begin{enumerate}
\item From Table \ref{table1}, it is apparent that the $V/I$ Stokes parameter appears as a descriptor of imaginary and real WVA, respectively, of linear ($x-y$), and circular diattenuation effects. Thus, the real and imaginary WVA schemes of measurement themselves make these two effects perfectly discernible from each other.
\item From Table \ref{table1}, it is also evident that the $V/I$ parameter appears as a descriptor of imaginary WVA of linear ($x-y$) diattenuation and circular retardance effects. In this scenario, these two effects can easily be discerned and separately quantified by performing WVA measurements with two different input polarization states, as is also evident from Table \ref{table1}. 
\end{enumerate}

The above argument is valid for any other polarization anisotropy effects. Hence, all the polarization effects can be separately extracted and quantified by performing real and imaginary WVA and/or with a clever choice of the input polarization states. 
\par
We note here that, as evident in any other WVA scheme, we have chosen the coupling parameter as the weak polarization anisotropy effect ($\alpha$) and achieved an amplification by a factor of $\cot{\epsilon}$. In essence, this implies that by applying this protocol of interferometric WVA, one can measure the enhanced Stokes parameters ($Q/I$, $U/I$, $V/I$) as $\sim \alpha \cot{\epsilon}$ and use that to quantify the corresponding anisotropy effect $\alpha$. In terms of sensitivity, this framework allows one to quantify anisotropy effects present in a sample that are $\epsilon$ times smaller than the sensitivity provided by a given Stokes polarimeter. Although we used a dc Stokes polarimeter which does not allow higher sensitivity, our proposal is generic in the sense that it is valid for polarimeters of higher sensitivity as well \cite{modak2021generalized}.

\begin{table*}[ht]
\caption{Real and imaginary WVA of all the polarization anisotropy effects. The four different polarization anisotropy effects (1st column), the corresponding input electric field and the electric field after experiencing a small anisotropy effect (2nd column), the Stokes vector elements carrying signature of real and imaginary WVA of respective anisotropy effects (3rd column). $E_0$ is polarization independent field amplitude. This table is adopted from \cite{modak2021generalized}.\\}
\label{table1}
\begin{tabular}{lccccr}
\centering
\multirow{2}{*}{Anisotropy effects} & \multicolumn{2}{c}{Electric fields}  & \multicolumn{2}{c}{Stokes vector elements}&
\\ \\ & \begin{tabular}[c]{@{}c@{}}$\frac{\vec{E_1}}{E_0}$ \end{tabular} & $\frac{\vec{E_2}}{E_0}$ & Real WVA  & Imaginary WVA  & \\ \\
Linear diattenuation\\ ($x -y$)& \begin{tabular}[c]{@{}c@{}}$\frac{\hat{x}+\hat{y}}{\sqrt{2}}$\end{tabular} & $\frac{e^\alpha\hat{x}+e^{-\alpha}\hat{y}}{\sqrt{e^{2\alpha}+e^{-2\alpha}}}$ & $\frac{Q}{I}$ & $\frac{V}{I}$ 
\\ \\Circular diattenuation & \begin{tabular}[c]{@{}c@{}}$\hat{x}=\frac{\hat{r}+\hat{l}}{\sqrt{2}}$\end{tabular} & $\frac{e^\alpha\hat{r}+e^{-\alpha}\hat{l}}{\sqrt{e^{2\alpha}+e^{-2\alpha}}}$ & $\frac{V}{I}$ & $\frac{U}{I}$ 
\\\\Linear retardance & \begin{tabular}[c]{@{}c@{}}$\frac{\hat{x}+\hat{y}}{\sqrt{2}}$\end{tabular} & $\frac{e^{i\alpha}\hat{x}+e^{-i\alpha}\hat{y}}{\sqrt{2}}$ & $\frac{V}{I}$ & $\frac{Q}{I}$ 
\\\\Circular retardance & \begin{tabular}[c]{@{}c@{}}$\hat{x}=\frac{\hat{r}+\hat{l}}{\sqrt{2}}$\end{tabular} & $\cos{\alpha}\hat{x}+\sin{\alpha}\hat{y}$ & $\frac{U}{I}$ & $\frac{V}{I}$ 
\end{tabular}
\end{table*}

Hence, the proposed weak value polarimeter is able to enhance small polarization anisotropy effects and lays out a way to quantify those effects. However, most naturally evolving substances come with more than one anisotropy effects that are often exhibited simultaneously in space \cite{collett1993polarized,gupta2015wave,perez2016polarized,ghosh2011tissue,azzam2016stokes}. Even for a single anisotropic element, these do not appear in any preferential order in the most realistic scenario \cite{collett1993polarized,gupta2015wave,perez2016polarized,ghosh2011tissue,azzam2016stokes}. This practical scenario naturally stimulates one to extend the proposed technique to invoke multiple anisotropy effects simultaneously.
\par
Following similar interferometric approach \cite{modak2021generalized}, we can write the Hamiltonian and hence, the corresponding unitary evolution for weak interaction between the system and pointer in simultaneous presence of multiple tiny polarization effects $\alpha$, $\beta$, $\gamma$, \ldots ($\alpha, \beta, \gamma, \ldots\rightarrow 0$) as
\begin{equation}
    U=e^{-i (\alpha J_{\alpha}+\beta J_{\beta}+\gamma J_{\gamma}+\ldots)\hat{A}}
    \label{eq1}
\end{equation}
with $J_{\alpha}$, $J_{\beta}$, $J_{\gamma}$, \ldots be the generators respectively which act as the measuring pointer in our interferometric system. $\hat{A}=\begin{pmatrix}
    1 & 0\\
    0 & -1
    \end{pmatrix}$, written in the path basis \cite{modak2021generalized}, essentially implying that anisotropy effects $\alpha$, $\beta$, $\gamma$, \ldots are imposed oppositely in either arm of the interferometer \cite{modak2021generalized}. Considering the anisotropy effects to be small enough, the first-order approximation leads to the final state $\ket{\phi_f}$ after post-selection in the following form \cite{modak2021generalized}.
\begin{equation}
    \ket{\phi_f}\sim[\mathbb{I}-i A_w (\alpha J_{\alpha}+ \beta J_{\beta} + \gamma J_{\gamma} +\ldots)]\ket{\phi_i}
    \label{eq2}
\end{equation}
Here, $A_w=\frac{\mel{\psi_f}{\hat{A}}{\psi_i}}{\braket{\psi_f}{\psi_i}}$ is the conventionally defined as the weak value (WV) of the operator $\hat{A}$, $\ket{\psi_i}$, $\ket{\psi_f}$ are the pre and post-selected states of the system, $\ket{\phi_i}$ is input pointer state \cite{modak2021generalized}. Hence, Eq. \eqref{eq2} indicates an enhancement of the simultaneously given anisotropy parameters upon a careful choice of pre and post-selection, i.e., when post-selection is nearly orthogonal to pre-selection. Near orthogonal pre and post-selection implies near destructive interference of the slightly shifted pointer states as mentioned above, and thus, the small overlap between pre and post-selection is equal to the $\epsilon_{a\slash p}$ parameter. 
\subsection{Joint WVA of circular retardance and diattenuation effect}
\label{sec2.1}
For a physically tangible situation, we now take an example of simultaneous presence of circular retardance $\alpha$ and circular diattenuation $\beta$ with the respective generators \(J_{\alpha}\) and \(J_{\beta}\) which are commonly found in chiral substances \cite{collett1993polarized,gupta2015wave,perez2016polarized,ghosh2011tissue,azzam2016stokes}. The corresponding unitary evolution $U=e^{-i (\alpha J_{\alpha} +\beta J_{\beta})\hat{A}}$ \cite{modak2021generalized}. The generators of the anisotropy effects are expressed in terms of Pauli matrces $\sigma_j$ ($j=1,2,3$), $J_{\alpha}=\frac{1}{2} \sigma_2$ and $J_{\beta}=\frac{i}{2} \sigma_2$ \cite{han1997stokes}. Taking $\ket{\phi_i}=\begin{pmatrix}1\\0\end{pmatrix}$, the final state of pointer for real ($\ket{\phi_f^r}$) and imaginary ($\ket{\phi_f^i}$) WVA is obtained after successful post-selection $\ket{\psi_f}$.

\begin{eqnarray}
    \ket{\phi_f^r}&\sim& \begin{pmatrix}
    1 \\
    \frac{\alpha+i\beta}{2} \cot{\epsilon_a}
    \end{pmatrix} \label{eq3}\\
    \ket{\phi_f^i}&\sim& \begin{pmatrix}
    1 \\
    \frac{-i\alpha+\beta}{2} \cot{\epsilon_p}
    \end{pmatrix} \label{eq4}
      \end{eqnarray}
      
where \begin{subequations}
\begin{equation}
    \ket{\psi_i}=\frac{1}{\sqrt{2}}[\ket{1}+\ket{2}]
    \label{eq3a}
\end{equation}
 \begin{equation}
   \ket{\psi_f}=\frac{1}{\sqrt{1+a^2}}[e^{- i\epsilon_p}\ket{1}-a e^{+ i\epsilon_p}\ket{2}]\label{eq3b}
   \end{equation}
   \label{eq5}
\end{subequations}
In our interferometric WVA approach, Eq. \eqref{eq3} and \eqref{eq4} represent the Jones vector. A straightforward extension from Jones to Stokes vector indicates the characteristic Stokes vector elements displaying the enhancement of small anisotropy effects $\alpha$ and $\beta$ owing to real and imaginary WVA. Mathematical manipulation indicates that for real WVA, $U/I\sim \alpha \cot{\epsilon_a}$ and $V/I\sim \beta \cot{\epsilon_a}$ demonstrating the real WVA of circular retardance $\alpha$ and circular diattenuation $\beta$ respectively. Similarly, for imaginary WVA, $V/I\sim \alpha \cot{\epsilon_p}$ and $U/I\sim \beta \cot{\epsilon_p}$ respectively. Hence, for the simultaneous presence of both the anisotropy effects, real and imaginary WVA enhances both effects, and the corresponding enhancements are reflected in the same characteristic Stokes vector elements listed in Table \ref{table1}.

\begin{figure}[hbt!]
\centering
\includegraphics[width=0.8\linewidth]{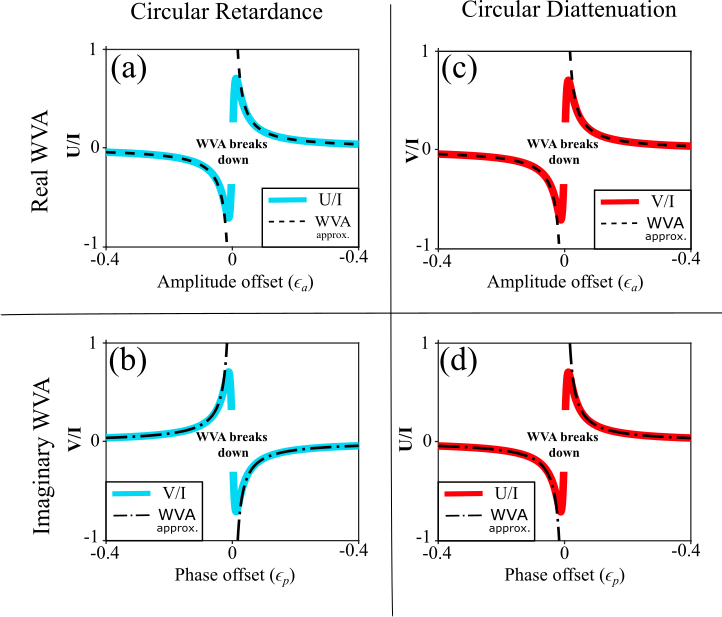}
\caption{Real and imaginary WVA of circular retardance ($\alpha=0.017$) ((a) and (b)) and circular diattenuation ($\beta=0.017$) ((c) and (d)) effect. (a) The variation of the third Stokes parameter $U/I$ is plotted with changing amplitude offset parameter $\epsilon_a$ as blue soild line. Corresponding linear response regime \cite{aharonov1988result,dressel2014colloquium,kofman2012nonperturbative,modak2020weak} approximation of WVA is given by black dashed line. (b) The variation of the fourth Stokes parameter $V/I$ is plotted as blue solid line with changing phase offset parameter $\epsilon_p$. Black dash-dotted line demonstrates the corresponding WVA approximation. (c) and (d) Similar variation of $V/I$ (red solid line) and $U/I$ (red slid line) is depicted with their WVA approximation $\sim \beta \cot{\epsilon_{a\ \text{and}\ p}}$ (black dashed line and black dash-dotted line respectively) as a function of $\epsilon_a$ and $\epsilon_p$ respectively. Beyond the linear response regime, the WVA breaks down as depicted.}
\label{fig2}
\end{figure}
Fig. \ref{fig2} demonstrates the enhancements of real and imaginary WVA of simultaneous presence of circular retardance $\alpha$ ($=0.017$) and circular diattenuation $\beta$ ($=0.017$). In Fig \ref{fig2} (a), the variation of $U/I$ with amplitude offset parameter $\epsilon_a$ is plotted as blue solid line. Corresponding WVA approximation ($\sim\alpha \cot{\epsilon_a}$) depicted as black dashed line shows a good agreement with the enhancement of $U/I$. Like other WVA protocols, here also the amplification of $U/I$ breaks down beyond linear response regime \cite{dressel2014colloquium,kofman2012nonperturbative,modak2020weak}, $\epsilon_a\rightarrow 0$. The imaginary WVA of circular retardance effect is demonstrated as the variation of $V/I$ with changing phase offset parameter $\epsilon_p$ by the solid blue line in Fig \ref{fig2} (b). Corresponding WVA approximation ($\sim\alpha \cot{\epsilon_p}$), depicted as a black dash-dotted line, shows a good agreement with the enhancement of $U/I$. Here also, WVA breaks down at $\epsilon_p\rightarrow 0$. The real and imaginary WVA of circular diattenuation effect $\beta=0.017$ are demonstrated in \ref{fig2} (c) and (d) respectively. The WVA approximation $\sim \beta\cot{\epsilon_{a\slash p}}$ shows good agreement with the characteristic Stokes vector elements $V/I$ and $U/I$ respectively. Fig. \ref{fig2} conclusively demonstrates that simultaneous presence of more than one weak anisotropy effects (circular retardance and circular diattenuation in case of Fig. \ref{fig2}) in sample can be individually enhanced. Thus, one can detect and quantify all the minute polarization anisotropy effects separately even though these appear simultaneously.

\subsection{Classical interference approach encompassing differential Jones Matrix}
\label{sec2.2}
Up till now, it has been shown that simultaneously appearing multiple small polarization anisotropy effects can be modeled by a unitary evolution operator corresponding to the weak interaction Hamiltonian. Now, we complement the above weak measurement formalism dealing with simultaneous circular retardance $\alpha$ and circular diattenuation $\beta$ effects with its corresponding classical field-based analog \cite{modak2021generalized, guchhait2020natural}. For describing such a sample with multiple anisotropy effects, it is required to define the differential Jones matrix $\boldsymbol{N}$ at every point, whereas the entire sample is represented by the exponential Jones matrix $\boldsymbol{M}$ \cite{Jones:48,Arteaga:09}. 
The relation between $\boldsymbol{M}$ and $\boldsymbol{N}$ is governed by the differential equation $\frac{d\boldsymbol{M}}{dz} = \boldsymbol{N}\boldsymbol{M}$ which gives the exponential form of the actual Jones matrix $\boldsymbol{M}$ for a homogeneous sample \cite{Jones:48}. The simultaneous presence of aforementioned anisotropy effects can be modeled as follows. 
\begin{equation}
\boldsymbol{M} = \exp[(\boldsymbol{N_1} + \boldsymbol{N_2})z] = \exp[i(\omega z J_{\alpha} + \delta z J_{\beta})]=\exp [i(J_{\alpha} \alpha + J_{\beta} \beta)]
   \label{eq6}
\end{equation}
where $\boldsymbol{N_1} = i \omega J_{\alpha}$ is the differential Jones matrix for circular  birefringence $\omega$, $\boldsymbol{N_2} = i \delta J_{\beta}$ is the differential Jones matrix for circular dichroism $\delta$ and $z$ is the path length traversed by light in the sample. Circular retardance $\alpha$ and circular diattenuation $\beta$ are thus the accumulated circular birefringence and circular dichroism effect respectively as light traverses through the sample. 
\begin{equation}
    \alpha=\omega z; \ \beta=\delta z
\end{equation}
\par
Now, we consider near destructive interference of two paths of an interferometer having slightly different polarizations of light \cite{modak2021generalized,guchhait2020natural} with electric field vectors
\begin{equation}
    \vec{E_1} = E_0\exp [i(J_{\alpha} \alpha + J_{\beta} \beta)]\hat{x};\ \vec{E_2} = E_0\exp [-i(J_{\alpha} \alpha + J_{\beta} \beta)]\hat{x}
    \label{eq7}
\end{equation}
where $E_0$ is the polarization independent field amplitude. As mentioned above, the real and imaginary WVA of $\alpha$, and $\beta$ can be obtained by nearly destructive interference of $\vec{E_1}$ and $\vec{E_2}$ with small amplitude offset ($\pm2\epsilon_a$) and small phase offset ($\pm2\epsilon_p$), respectively. The corresponding expressions for the resultant electric field for the real and the imaginary WVAs are
\begin{equation}
    (\cos{\epsilon_a}\mp\sin{\epsilon_a})\vec{E_1}-(\cos{\epsilon_a}\pm\sin{\epsilon_a})\vec{E_2};\ (\cos{\epsilon_p}\mp i\sin{\epsilon_p})\vec{E_1}-(\cos{\epsilon_p}\pm i\sin{\epsilon_p})\vec{E_2}
    \label{eq8}
\end{equation}
In the weak interaction limit $\alpha,\ \beta\rightarrow0$ \cite{aharonov1988result,guchhait2020natural,modak2021generalized}, the calculated Stokes vector from Eq. \eqref{eq7} takes the following form for real WVA.
\begin{equation}
    (U/I)_d\sim \alpha \cot{\epsilon_a};\ (V/I)_d\sim \beta \cot{\epsilon_a}
    \label{eq9}
\end{equation}
\begin{figure}[hbt!]
\centering
\includegraphics[width=0.8\linewidth]{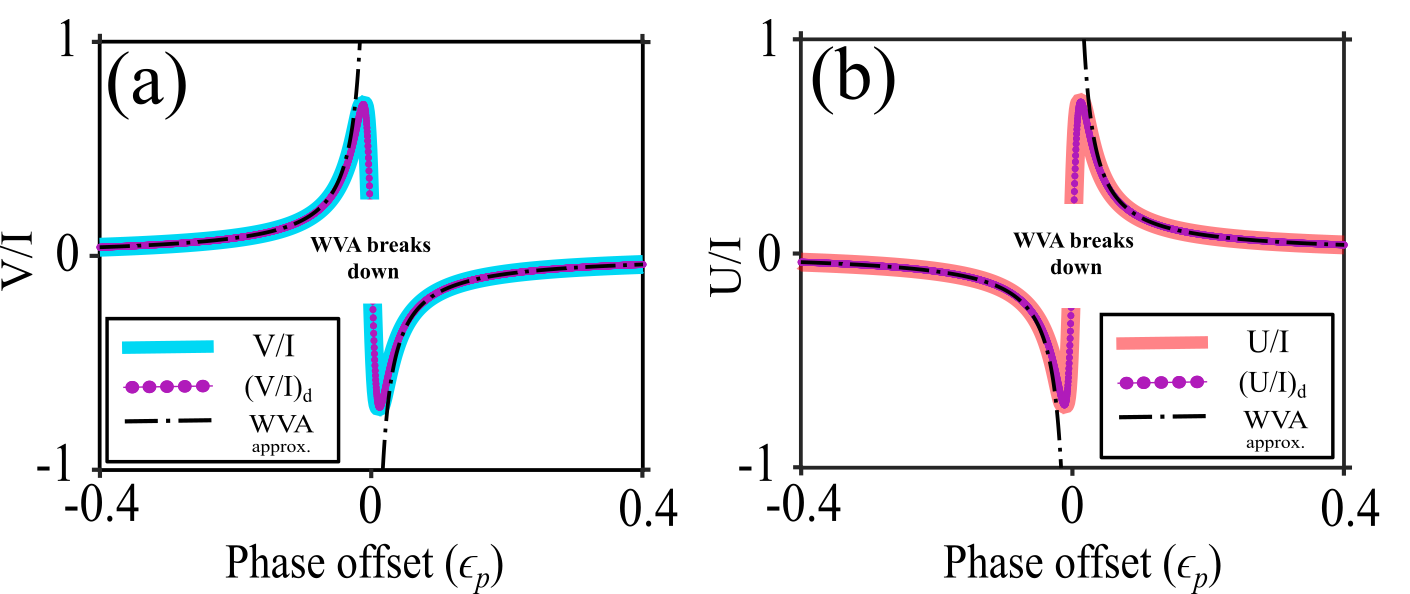}
\caption{Comparison of post-selected weak measurement approach and classical field based formalism through imaginary WVA of simultaneous (a) circular retardance ($\alpha=0.017$) (b) and circular diattenuation ($\beta=0.017$) effect. (a) The variation of the fourth Stokes parameter $V/I$ is plotted with changing phase offset parameter $\epsilon_p$ using Eq. \eqref{eq4} as blue solid line. Purple solid circled line demonstrates the same (variation of $(V/I)_d$) using Eq. \eqref{eq8}. Corresponding linear response regime approximation of WVA \cite{dressel2014colloquium,kofman2012nonperturbative,modak2020weak} ($\sim \alpha \cot{\epsilon_{p}}$) is given by black dash-dotted line. (b) Using Eq. \eqref{eq3} the similar variation of the third Stokes parameter $U/I$ is plotted as red solid line with changing phase offset parameter $\epsilon_p$. Purple solid circled line demonstrates the same (variation of $(U/I)_d$) using Eq. \eqref{eq8}. Black dash-dotted line demonstrates the corresponding WVA approximation $\sim \beta \cot{\epsilon_{p}}$. Beyond the linear response regime \cite{dressel2014colloquium,kofman2012nonperturbative,modak2020weak}, the WVA breaks down as depicted.}
\label{fig3}
\end{figure}
`d' stands for the differential Jones matrix formalism. This enables the characteristic Stokes vector elements ($U/I$ and $V/I$) to detect the enhanced anisotropy effects ($\alpha$ and $\beta$). Similarly for imaginary WVA, the characteristic Stokes vector elemnents appear as follows.
\begin{equation}
    (V/I)_d\sim \alpha \cot{\epsilon_p};\ (U/I)_d\sim \beta \cot{\epsilon_p}
    \label{eq10}
\end{equation}
 Thus, both the post-selected weak measurement approach and classical field interference approach yield similar real and imaginary WVA of both the anisotropy effects present simultaneously in a sample. More importantly, this interferometric WVA protocol provides a platform to measure the differential Jones matrix of that anisotropic sample of interest through the enhanced estimation of $\alpha$ and $\beta$.
\par
The way these two formalism yield similar results on performing simulation is illustrated in the Fig. \ref{fig3} for simultaneously present circular retardance $\alpha$ and circular diattenuation $\beta$ effect. In Fig. \ref{fig3} (a), the fourth Stokes parameter $V/I$ relevant for the imaginary WVA of circular retardance ($\alpha=0.017$) effect is plotted with changing phase offset parameter $\epsilon_p$. The corresponding $(V/I)_d$ from field interfernce-based approach is depicted as purple solid circled line using Eq. \eqref{eq8} and the fields $\vec{E_1}$ and $\vec{E_2}$ are taken from the Eq. \eqref{eq7}. As apparent, this two approach coincides exactly. The black dash-dotted line demonstrates the corresponding WVA approximation $\sim \alpha \cot{\epsilon_p}$. The similar plot for the imaginary WVA of circular diattenuation ($\beta=0.017$) effect is provided in Fig. \ref{fig3} (b). It is to be observed that the exponential matrix $M$ (Eq. \eqref{eq6}) has almost similar mathematical form as the joint weak measurement evolution operator $\eqref{eq1}$. This way, the interferometric WVA framework provides an efficient platform to study the differential Jones characterization of a sample. An extension of the interferometric WVA formalism to encompass depolarization effect is also warranted. In such case, the correspondence to this formalism will be through the differential Mueller matrix \cite{ossikovski2011differential} of a general depolarizing anisotropic sample. We are currently expanding our investigation in this domain including its experimental realization.



\section{Conclusion and discussions}
\label{sec3}
We have extended the recently proposed interferometric WVA technique to amplify tiny polarization anisotropy effects in a more practical scenario dealing with simultaneous exhibition of multiple tiny polarization anisotropy effects.  Polarization is used as the pointer here, whereas, path degree of freedom of the interferometer is considered as the system. Near destructive interference of the paths of the interferometer gives rise to the amplification of the anisotropy effects in the relevant Stokes vector elements. Although, like other WVA protocols, the enhancement of tiny polarization anisotropy effects comes at the cost of total intensity \cite{ferrie2014weak,harris2017weak}, but careful tuning of pre and post-selection enables to detect all possible polarization anisotropy effects, even if these appear simultaneously. Under linear approximation of WVA \cite{aharonov1988result,kofman2012nonperturbative,dressel2014classical,modak2020weak}, all the present anisotropy effects leave their signature enhancements in the characteristic Stokes vector elements which can be quantified under clever choice of pre and post-selection. From the practical point of view, it is important to quantify simultaneously exhibiting multiple polarimetry effects as in most of the realistic scenarios, these effects appear jointly in space even for a single anisotropic unit. This Interferometric WVA technique is particularly suited for such measurements where one deals with very weak polarization signals. For non-depolarizing samples, this approach provides a direct measurement of the corresponding differential Jones matrices that encodes simultaneous multiple polarization anisotropy effects. Thus this proposed protocol is a new step in the old collaboration between polarization measurement and WVA technique \cite{ritchie1991realization,hosten2008observation,dixon2009ultrasensitive,PhysRevA.89.053816,salvail2013full,rhee2015chiroptical,luo2019simultaneously,xu2019multifunctional,pryde2005measurement,iinuma2011weak,de2014ultrasmall,zhu2019single,dziewior2019universality,PhysRevLett.91.180402,PhysRevLett.93.203902}. An extension towards depolarizing samples will need further studies. Additionally, this interferometric setting produces a platform to study joint weak value \cite{resch2004extracting,strubi2013measuring,kobayashi2012extracting,kumari2017joint,lundeen2005practical,lundeen2009experimental} incorporating the 2nd order terms of Eq. \eqref{eq2}, choosing the anisotropy effect judiciously, and tailoring the interferometer wisely.

\bibliographystyle{ieeetr}
\bibliography{ref}

\end{document}